\documentclass[twocolumn,showpacs,preprintnumbers,amsmath,amssymb]{revtex4}

\usepackage{graphicx}% Include figure files
\usepackage{dcolumn}% Align table columns on decimal point
\usepackage{bm}% bold math
\usepackage{tabls}
\usepackage[english]{babel}
\usepackage{multirow}
\usepackage[latin1]{inputenc}

\begin{document}

\preprint{APS/123-QED}

\title{Robustness of the European power grids under intentional attack}

\author{Ricard V. Sol\'e$^{1,2}$, Mart\'i Rosas-Casals$^{1,3}$, Bernat Corominas-Murtra$^1$ and Sergi Valverde$^1$}

\affiliation{
$^1$ ICREA-Complex Systems Lab, Universitat Pompeu Fabra, Dr. Aiguader 80, 08003 Barcelona, Spain\\
$^2$Santa Fe Institute, 1399 Hyde Park Road, New Mexico 87501, USA
$^3$Catedra UNESCO de Sostenibilitat, Universitat Politecnica de Catalunya, EUETIT-Campus Terrassa, 
08222 Barcelona, Spain}

\begin{abstract}
The power grid  defines one of the most  important technological networks of our times and sustains
our complex society.  It has evolved for more than a century into an extremely huge and seemingly robust and well understood system. But it becomes extremely fragile as well,  when unexpected, usually minimal, failures turn into unknown dynamical behaviours leading, for example, to sudden and massive blackouts. Here we  explore the fragility of the European power grid 
under the effect of selective node removal. A mean field analysis of fragility against attacks is presented together
with the observed patterns. Deviations from the theoretical conditions for network percolation (and fragmentation) under attacks are analysed and correlated with non topological reliability measures.
\end{abstract}

% PACS, the Physics and Astronomy Classification Scheme.
%\keywords{Suggested keywords}
%Use showkeys class option if keyword display desired
\maketitle

\section{Introduction}%%%%%%%%%%%%%%%%%%%%%%%%%%%%%%%%%%%%

The power grid defines,  together with transportation networks and the
Internet, the  most important class  of human-based web.   It allows
the success of  advanced economies based on electrical  power but it
also illustrates  the limitations imposed by environmental concerns, together with economic 
and demographic growth:  the power grid reaches its  limits with an ever growing demand
\cite{EUgreenpaper}. A direct consequence of  this situation is the fragility of this
energy infrastructure, as  manifested in terms of sudden  blackouts  and large  scale cascading failures, mostly caused  by localized, small scale failures, ocurring at an increasing frequency \cite{TheGrid, LightsOut}.

The fragility of the power grid is an example of a generalized feature
of most complex  networks, from the Internet to  the genome \cite{BAreview, Bornholdt, SIAMreview, Dorogovtsev,LATORAreview}.
Specifically, real networks are  often characterized by a considerable
resilience against  random removal or failure of  individual units but
experience important  shortcomings when the  highly connected elements
are  the target  of  the  removal. Such  directed  {\em attacks}  have
dramatic   structural   effects, typically   leading   to   network
fragmentation \cite{AlbertGrid, MotterLai, Carreras, Motter}. This 
behaviour has been studied for  skewed
 power-law  distributions  of links, which are found in  many
small-world networks \cite{AlbertTolerance, CrucittiTolerance}. But recent studies
 have shown  that similar responses are not unique to
small-world,  scale-free networks:  power  grids, having  less skewed
exponential  degree   distributions  and  often   without  small-world
topology,  display  similar  patterns  of response  to  node  loss.
 \cite{Rosas2007}

An additional  feature  of   the  power  grid  is   its  spatial
structure. The  geographic character of this network  implies
 that a number of  constraints are  expected to be  at work. Other  well known
spatially extended nets include  the Internet \cite{Internet}, street
networks    \cite{Streets},     railroad    and    subway    networks
\cite{railsubway},   ant  galleries  \cite{Ants}, electric  circuits
\cite{Circuits} or cortical graphs \cite{Cortical}.
 
One fundamental aspect concerning the analysis of complex networks 
is the increasing evidence of mutual influence between dynamical behavior and topological structure. 
The topology of human contact networks, for example, determines the emergence of epidemics \cite{EpidemicSpreading}; similarly, the correct dynamics in cellular networks are rooted in the 
topology of the regulatory networks \cite{AlbertOthmer, Maayan}. Here we present
 evidence of a plausible relation between topological and non topological reliability measures
for the power grid, suggesting that topology might be capturing the robustness (or fragility) of the real system, when dynamics are at work. This evidence has been obtained analysing the resilience of 33 different power grids: (a) the 23 different EU countries, (b) 4 geographically related zones (Iberian Peninsula, Ireland as island, England as island and United Kingdom and Ireland as a whole), (c) 4 traditionally united or separated regions (former Yugoslavia, Czechoslovaquia and Federal and Democratic Republics of Germany), (d) continental Europe and (e) continental Europe plus United Kingdom and Ireland. 

The  paper is
organized as  follows. In  section II the  data set on  European power
grids is presented and their basic topological features summarized. In
section III we present both  analytical and numerical estimations of the
boundaries for network collapse  under attack, using a mean field theoretical approach.
Two classes of networks are shown  to be present. In  section IV,  evidence for correlation
between these  two classes and non topological reliability indexes
is shown to exist. In section V we summarize our findings and outline
their implications.

\section{Power grid data sets}%%%%%%%%%%%%%%%%%%%%%%%%%%%%%%%%%%%%

\begin{figure*}
\includegraphics[width=1.0\textwidth]{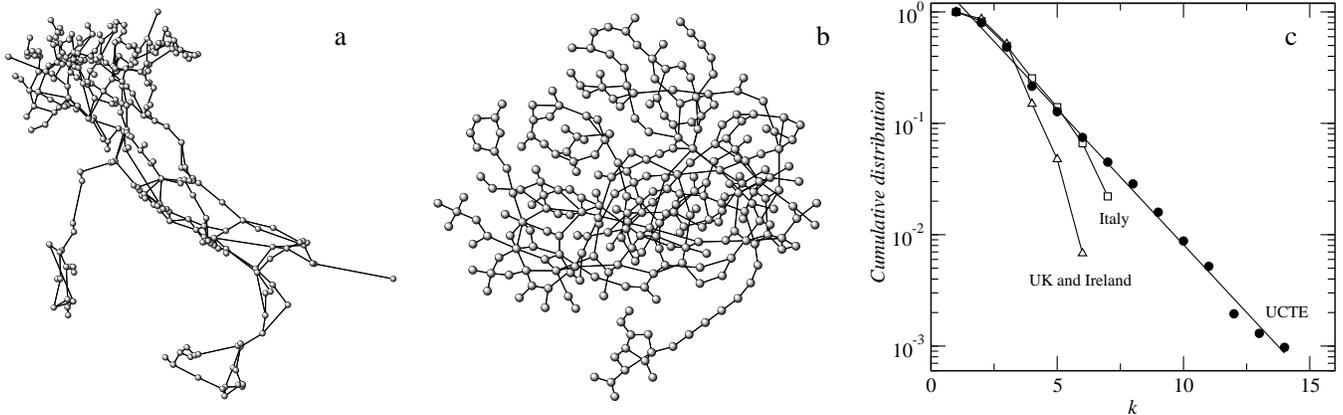}
\caption{Power  grids define  a spatial,  typically planar  graph with
nodes  including generators, transformers and substations.  Here we  show  (a) the
geographical  and (b)  the topological
  organization of  the Italian
power grid.  These webs are homogeneous, having  an exponential degree
distribution, $P(k)=\exp(-k/\gamma)/\gamma$
 as shown in (c).}
\label{Italia}
\end{figure*}

Europe's electricity transport network is nowadays the ensemble of more than twenty different national power grids
coordinated, at its higher level, by the Union for the Co-ordination of Transmission Electricity, UCTE (http://www.ucte.org).
The distribution and location of transmission lines, plants, stations, etc. can be found in the last version (July 2007) of the UCTE Map. The different data sets analyzed here have been obtained after introducing the topological values (i.e. geographical positions and longitudes) of more than 3,000 generators and substations (nodes) and 200,000 km of transmission lines (edges) in a geographical information system (GIS). The national power grid for every country  or region has been obtained from a typical GIS query: the selection of the part of the UCTE's network constrained by every country's frontier. The  power  grid  can  then be formally  described in  terms  of  a  graph $\Omega=(V,E)$.  Here  $V=\{ v_i  \}$ indicates the  set of  $N$ nodes (transformers, substations or  generators in  our  context).  Figure  \ref{Italia} 
shows an example of such graphs with its geographical (a) and topological (b) structures, respectively. 
These  nodes can  be connected,  and $E=\{e_{ij}\}$  indicates the  set  of actual
links  between  pairs  of  nodes.  Specifically,  $e_{ij}=\{v_i,v_j\}$
indicates that  energy is being  transported between the nodes  in the
pair $\{v_i,v_j\}$.  Our system can  be analyzed at two  main levels:
the whole power grid  $\Omega_{EU}$ including all countries within the
EU and at the country  level. If $\Omega_k$ indicates the $k$-th power
grid of one of the $n=33$ countries and regions involved, we have $\Omega_{EU} = \bigcup_{k=1}^n \Omega_k$.

The global organization of  these webs has been previously analyzed
\cite{Rosas2007},   revealing  a  very   interesting  set   of  common
regularities: (a) most of them are small worlds (i.e. very short  path lengths are
tipically present) and the larger webs display clustering coefficients
much  larger  than expected  from  a  random  version of  the  network
analysed; (b) they are very sparse, meaning that the average number of
links  is such  that $\langle  k \rangle  \ll N$,  with an  average of
$\langle  k \rangle=2.8$  over all  the webs  available (see Table I); (c)  the link
distribution is  exponential: the probability of having  a node linked
to  $k$ other  nodes  is $P(k)=\exp(-k/\gamma)/\gamma$  (Fig. \ref{Italia}, c) and (d)
these networks  are weakly or  no correlated. This exponential distribution is thus characterized  by the  constant $\gamma$  which actually  corresponds  to  the  average  degree (i.e., $\langle  k\rangle=\gamma$).

Correlations   were  measured  using   the  average   nearest  neighbor
connectivity of a node with the degree $k$, i. e. the average $\langle k_{nn} \rangle = \sum_{k'} k' P(k'\vert k)$
where  $P(k'\vert  k)$ is  the  conditional  probability  that a  link
belonging  to a  node  with connectivity  $k$  points to  a node  with
connectivity $k'$  \cite{Satorras2001}. For  these webs, it  was found
that  $\langle k_{nn}  \rangle \approx$  constant, as  expected  if no
correlations  were present.  This is  a  very useful  property in  our
analysis, since  makes mean field  predictions valid in spite  that we
ignore  the planar  character of  these networks,  thus  replacing the
geographical pattern  by a topological one.

\section{Attacks in exponential networks: mean field theory}%%%%%%%%%%%%%%%%%%%%%%

\begin{figure}
\includegraphics[clip=true, width=.45 \textwidth]{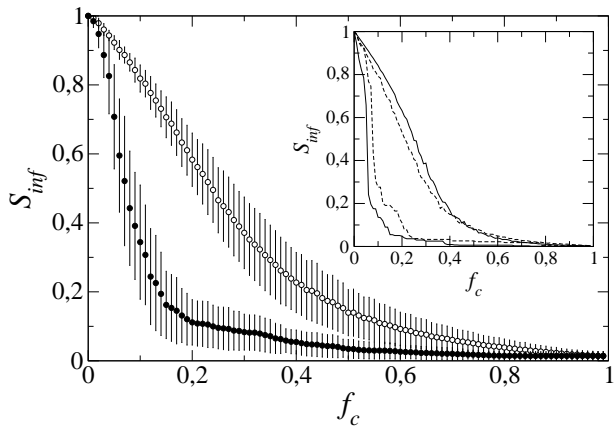}
\caption{Effects of  attacks and  failures on the  topology of  the EU
power grids. Static tolerance to random  (white circles)
and  selective (black circles) removal of a fraction $f$ of nodes, measured
by the relative size $S_{inf}$ of the largest connected component. Whiskers 
stand for the standard deviation.
Inset: Evolution of the static tolerance to random and selective node removal 
for Italy (dashed lines) and France (continous lines). Though in the case of random removal 
(failures) both networks exhibit a similar response, for the selective one (attacks), Italy behaves 
in a slightly stronger manner (i.e., for a fixed fraction of eliminated nodes, the relative size 
of the largest connected component in Italy remains always higher than that of France).}
\label{Attacks}
\end{figure}

In our  previous paper, we analysed  the effects of  both random and selective removal of nodes on
the EU grids \cite{Rosas2007}.  Nonetheless, in that paper we were mostly
interested in the  average behavior of the networks  analysed (see Figure \ref{Attacks}). Here we
want  to extend  these  results  to the  analysis  of the  differences
observed in EU power grids with the goal of interpreting the different
patterns exhibited compared to the predictions from mean field theory on 
intentional attacks.

In order to compute the effect  of random removal of nodes, we compute
the percolation condition for the graph assuming it is sparse
and  uncorrelated.  Let $f$  be the fraction of removed nodes  and $P(k)$  the link degree distribution  of our  graph. 
The damaged  graph   will  be   characterized  by  the   following  degree
distribution ${\mathbf P}(k)$:

\begin{equation}
{\mathbf P}(k)=\sum_{i\geqslant k}^{\infty}\binom{i}{k}f^{i-k}(1-f)^k P(k)
\end{equation}

Note that such an equation corresponds to the case when a fraction $f$
of nodes are  removed but it also holds when a  fraction $f$ of links
are  removed (or  lead  to  unoccupied sites).

In  order  to  study  percolation  properties,  we  use  the  standard
generating function  methodology.  The two  first generating functions
of the damaged graph are:

\begin{eqnarray}
F_0(x)=\sum_{k}^{\infty}P(k)(1-f)x^k \\
F_1(x)=\frac{1}{\langle k \rangle}\sum_k^{\infty}kP(k)(1-f)x^{k-1}
\label{binom}
\end{eqnarray}

\noindent
The  averages  (i.e.,  the   values  at  $x=1$)   are $F_0(1)=F_1(1)=1-f$   respectively.
Here $F_0(1)$ is the fraction of nodes from the
original graph belonging to the damaged graph.  Similarly, $F_1(1)$
is the  relation among $\langle k  \rangle$ and the  average number of
nodes from  $V$ that can be reached  after deleting a  fraction $f$  of nodes.
The generating  function for  the size of  the components, other  than the
giant one, which can be reached from a randomly choosen node is:

\begin{equation}
H_1(x)=f+xF_1(H_1(x))
\label{H1}
\end{equation}

\noindent
And the generating  function for the size of the  component to which a
randomly choosen node belongs to is \cite{Callaway}:

\begin{equation}
H_0(x)=f+xF_0(H_1(x))
\label{H0}
\end{equation}

\noindent
Thus, the average  component size, other than the  giant component will be:

\begin{equation}
\langle s \rangle=H'_0(1)=1-f+F'_0(1)\times H'_1(1)
\label{s}
\end{equation}

\noindent
After  some algebra,  we see  that this  leads to  a  singularity when
$F'_1(1)=1$.  To ensure  the  percolation of  the  damaged graph,  the
following inequality has to hold:

\begin{equation}
\sum_k k(k-2)P(k)>\sum_k k(k-1)f P(k)
\label{Percol1}
\end{equation}

\noindent
The above expression can be expressed as:

\begin{equation}
\langle k^2 \rangle -2\langle k \rangle >f(\langle k^2 \rangle -\langle k\rangle)
\label{Percol2}
\end{equation}

\noindent
which leads to a critical probability of node removal $f_c$ given by:

\begin{equation}
f_c = 1-{1 \over \kappa_0 - 1}
\label{fc}
\end{equation}

\noindent
where $\kappa_0=\langle k^2 \rangle/\langle  k \rangle$.  In our case,
we have  an analytic  estimate $\kappa_0=2\gamma$.  Using  the average
value $\langle  \gamma \rangle =1.9$, we obtain  a predicted critical
probability $f_c=0.61$.

Although  random  removal is  an  interesting  scenario, it  considers
chance events that are not correlated to network structure. Intentional
attacks strongly deviate from random failures: even a small fraction of
removed nodes having large degrees has dramatic consequences. In order
to predict the effects of  such directed attacks on network structure, the  critical probability  associated  to network
breakdown can be computed.  Here  we  follow  the  formalism developed  by  Cohen  et
al. \cite{Cohen2000}.   Roughly speaking, this  formalism enables
us to  {\em translate} an  intentional attack into an  equivalent random
failure  and   study    the problem in terms of standard   percolation    using
equation (\ref{fc}). When the selective removal  of the most connected nodes
is considered, a  fraction of  order ${\cal  O}(1/N)$ is removed by
eliminating elements  with a  degree larger than  a given  $k=K$. This
upper cutoff is then easily computed from the continuous approximation:

\begin{equation}
\sum_K^{\infty}P(k)\approx \int_K^{\infty} {1 \over \gamma} e^{-k/\gamma} dk = {1 \over N}
\end{equation}

\noindent
and  the  new  cutoff  $\tilde{K}$  can be  obtained  (again  under  a
continuous approximation) from:

\begin{equation}
\int_K^{\tilde{K}} {1 \over \gamma} e^{-k/\gamma} dk = \int_K^{\infty} {1 \over \gamma} e^{-k/\gamma} dk - {1 \over N} = p
\end{equation}

\noindent
which gives (assuming $K$ large enough) a new cutoff:

\begin{equation}
\tilde{K} = - \gamma \ln p
\label{K}
\end{equation}

Following \cite{Cohen2000}, we translate the problem of
intentional  attack  to an  equivalent  random  failure problem.   The
removal of a  fraction $f$ of nodes  with the highest degree is then equivalent to
the random removal of those links connecting the remaining nodes to those already removed.
Thus, the probability that a specific link leads to a deleted node will be given by:

\begin{equation}
\tilde{p} = \int_K^{\tilde{K}} {k P(k) \over \langle k \rangle} dk
\end{equation}

\noindent
being  $\langle  k  \rangle$  the  average  degree  of  the
undamaged graph.  It is not difficult to show that this gives:

\begin{equation}
\tilde{p} = \left ( {\tilde{K} \over \gamma} + 1 \right ) e^{-\tilde{K}/\gamma}
\end{equation}

Using equation (\ref{K}) it is straightforward to see that:

\begin{equation}
\tilde{p} =( \ln p_c - 1 ) p_c
\end{equation}

\noindent
where  we  assume  that $K$  is  large  enough  to ignore  the  term
$\exp(-K/\gamma)$.   Thus, an equivalent  network with
maximal degree  $\tilde{K}$  has been built after a random  removal of
$\tilde{p}$  nodes due  to the  fact  that  the absence  of
correlations implies a  random failure of links. In order  to  obtain the
degree distribution  of the damaged  graph, such a failure
can  be introduced  into equation (\ref{binom}). But this will  be formally
equivalent  to  the removal  of  $\tilde{p}$  nodes.   Thus, to  study
stability   properties,  we  only   need  the   resulting  probability
$\tilde{p}$ to be introduced in the critical condition for percolation
(\ref{fc}).  Replacing $p_c=\tilde{p}$, we obtain:

\begin{equation}
1 +  ( \ln p_c - 1 ) p_c  = {1 \over 2 \gamma - 1}
\end{equation}

%%%%%%%%%%%%%%%%%%%%BARNATUS_DIXIT%%%%%%%%%%%%%%%%%%%%%%%%%%%%%

\begin{figure}
\includegraphics[width=.4\textwidth]{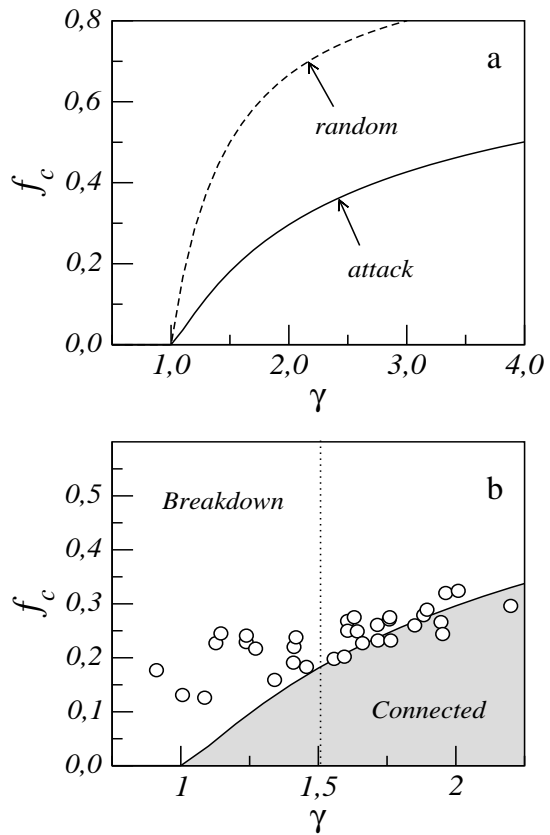}
\caption{(a) Phase  space for exponential  uncorrelated networks under
random removal  of nodes and directed attack  towards highly connected
vertices.  Here $\gamma$  is  the average  degree  of the  exponential
network and $f_c$ indicates the  fraction of removed nodes required in
order to  break the network into  many pieces. The upper  curve is the
critical  boundary for  network  percolation under  random removal  of
nodes.  Below it, a  network experiencing  such random  failures would
remain  connected (i.  e. with  a  giant component).  The lower  curve
corresponds to  the critical boundary  for attacks. In (b)  we display
the estimated  values of  $f_c(\gamma)$ for attacks  from the  thirty-three EU power
grids (circles) to be  compared with the mean  field prediction
(continuous line).  }
\label{Phase_space}
\end{figure}

\noindent
whose solutions  (for each fixed $\gamma$) provide  the conditions for
network percolation  under attacks. In Figure  \ref{Phase_space} and Table I, we show  the result of
our calculations. As  expected, a much lower value of  $f_c$ is required to
break a  power grid  network through intentional  attack.  

Now  we can compare  this mean  field prediction, evaluated as $f_c^{theor}$,  with available  data.  Using the
whole  dataset of EU  grids, we  can estimate  $f_c^{real}$ for  all EU
countries. The result are shown, for both $f_c$'s, in Figure \ref{Phase_space}, b. As we can see, there 
is a very good agreement (given their small size) between observed (real) and predicted (theoretical)  $f_c$  values, but some 
nontrivial deviations are also obvious.  We  can see that aproximately for
$\gamma>1.5$  the expected  $f_c$  values are  very  similar to  those
predicted by  theory. However, the power grids  having lower exponents
(when $\gamma < 1.5$) strongly deviate from the predicted values. These agreements and deviations are not due
to some simple statistical trait, such as network size. As indicated in Table I, very large power grids are in both sides (i.e,  the German and Italian grids are in the first group whereas the Spanish and French ones, belong to the second).

\begin{table*}
\begin{center}
%\begin{tabular*}{0.75\textwidth}{@{\extracolsep{\fill}} c | c | c c c | c c c | c | c | c }
%\begin{tabular}{l|c|ccc|ccc|c|c|c}
\begin{tabular}{p{2.5cm}|p{1.3cm}|p{1.4cm}p{1.4cm}p{1.4cm}|p{1.4cm}p{1.4cm}p{1.4cm}|p{1.4cm}|p{1.4cm}|p{1.4cm}}

\hline
\hline
\multicolumn{2}{c}{} & \multicolumn{3}{c}{Errors} & \multicolumn{3}{c}{Attacks}\\
Country & $\gamma$ & $f_c^{theor}$ & $f_c^{real}$ & $|\Delta f_c|$ & $f_c^{theor}$ & $f_c^{real}$ &
$|\Delta f_c|$ & $N$ & $L$ & $<k>$\\

\hline
{\emph{Belgium}}&1,005&0,011&0,395&0,384&0,010&0,131&0,121&53&58&2,18\\
{\emph{Holland}}&1,086&0,147&0,387&0,240&0,034&0,126&0,092&36&38&2,11\\
{\emph{Germany}}&1,237&0,322&0,565&0,243&0,097&0,229&0,132&445&560&2,51\\
{\emph{Italy}}&1,238&0,322&0,583&0,261&0,097&0,241&0,144&272&368&2,70\\
Austria&1,409&0,450&0,506&0,056&0,159&0,191&0,032&70&77&2,20\\
\emph{Rumania}&1,418&0,455&0,579&0,124&0,162&0,238&0,076&106&132&2,49\\
\emph{Greece}&1,457&0,477&0,492&0,015&0,174&0,183&0,009&27&33&2,44\\
Croatia&1,594&0,543&0,525&0,018&0,214&0,202&0,012&34&38&2,23\\
{\emph{Portugal}}&1,606&0,548&0,595&0,047&0,217&0,250&0,033&56&72&2,57\\
EU&1,630&0,557&0,629&0,072&0,223&0,275&0,052&2783&3762&2,70\\
\emph{Poland}&1,641&0,562&0,594&0,033&0,226&0,249&0,023&163&212&2,60\\
\emph{Slovakia}&1,660&0,569&0,563&0,006&0,231&0,227&0,004&43&52&2,41\\
{Bulgaria}&1,763&0,604&0,570&0,034&0,256&0,232&0,024&56&67&2,39\\
\emph{Switzerland}&1,850&0,629&0,610&0,020&0,275&0,260&0,015&147&186&2,53\\
\emph{Czech Republic}&1,883&0,638&0,634&0,004&0,281&0,279&0,003&70&88&2,51\\
{\emph{France}}&1,895&0,641&0,647&0,006&0,285&0,289&0,004&667&899&2,69\\
{\emph{Hungary}}&1,946&0,654&0,617&0,036&0,295&0,266&0,029&40&47&2,35\\
{Bosnia}&1,952&0,655&0,588&0,067&0,295&0,244&0,052&36&42&2,33\\
{\emph{Spain}}&2,008&0,668&0,689&0,020&0,307&0,324&0,017&474&669&2,82\\
{\emph{Serbia}}&2,199&0,705&0,655&0,051&0,339&0,296&0,054&65&81&2,49\\
\hline
\hline
\end{tabular}
\end{center}
\caption{A  summary of the  basic features  exhibited by  some of the European
power grids analyzed, ordered by  increasing $\gamma$, 
 the exponential degree distribution  exponent. The  critical 
  probability of  node removal $f_c$  is shown  for  both  cases, theoretical  and  real, and  random
(errors) and   selective (attacks)  removal of  nodes. The absolute difference
 $|\Delta f_c|$  between theoretical and observed critical
probability  diminishes  as $\gamma$  increases  in general  terms.
Number of  nodes $N$, number  of links $L$  and mean degree  $<k>$ are
also shown  as reference. 
%Countries in  boldface have  been used to evaluate subgraphs abundances.
Countries  in italics have been used to
evaluate reliability indexes.
 EU results (i.e., results for the  $\Omega_{EU}$ graph) are shown for comparative
purposes.}
\end{table*}

\section{Correlations with non-topological reliability measures}%%%%%%%%%%%%%%%%%%%%%

\begin{figure}
\includegraphics[width=.48\textwidth]{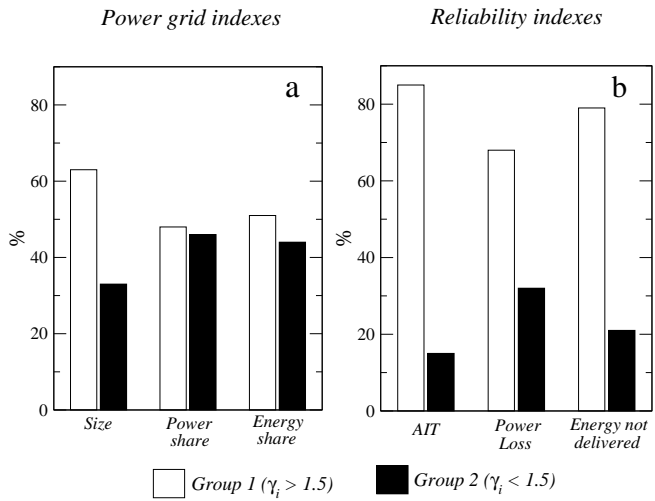}
\caption{Power grid indexes vs. reliability indexes. (a) Networks in group 1 (i.e., $\gamma > 1.5$ and $f_c \cong f_c,_p$), 
though represent two thirds of the UCTE size, share almost as much power and energy as networks in group 2 
(i.e., $\gamma < 1.5$ and $f_c > f_c,_p$). (b) Nonetheless, these same networks of group 1 acummulate more than
five times the average interruption time (AIT) of the latter, more than two times their power losses and almost
four times their undelivered energy.} 
\label{net_reliability}
\end{figure}

The reliability of a power grid evaluates its ability to continuously meet demand under major events 
like overloads, general failures, external impacts and alike. At the engineering level, and due to the different 
dimensions of service quality involved in a power grid (i.e., consumers, companies and regulators), reliability 
has been traditionally measured by different indexes as (a) the amount of energy not supplied, (b) the total loss of power or (c) the equivalent time of interruption, which measures the number and duration of interruptions 
experienced by customers \cite{CEER}. In this sense we would expect a correlation
between the critical percolation fraction $f_c$, the exponent that characterizes the grids' cumulative degree
distribution $\gamma$, and some of (if not all) these reliability indexes presented.

In order to explore the problem, three reliability indexes have been obtained from the UCTE monthly 
reliability measures \cite{UCTE}. They are related to four major events. Namely: overloads, 
general failures, external impacts and exceptional conditions, and finally other reasons (including unknown reasons). 
For every major 
event and transmission grid, the following indexes have been considered and normalized: 
(1) energy not supplied, normalized by the gross UCTE electricity consumption; (2) total loss of power, 
normalized by the UCTE peak load on the third Wednesday of December; and (3) equivalent time of interruption 
(also known as average interruption time or AIT), which is the ratio between the total energy not supplied and the average 
power demand per year, measured in minutes per year (normalized by definition).

In order to avoid statistical deviations due to the limited historical data available (UCTE monthly 
statistics have been published only from January 2002 onwards), we have devided UCTE networks in two groups.
Group 1 includes those countries whose critical breakdown probability $f_c^{real}$ agrees with that predicted $f_c^{theor}$
(i.e., countries with $\gamma > 1.5$). Group 2 includes those countries whose $f_c^{real}$ deviates positively from $f_c^{theor}$
 (i.e., countries with  $\gamma < 1.5$), with an expected more robust topology than that predicted. 

Figure \ref{net_reliability} gives the acummulated percentage values for the formerly presented reliability indexes 
and for each group of networks. As we can see, networks in group 1 (i.e., networks with $f_c^{real} \cong f_c^{theor}$)
represent 63\% of the whole UCTE nodes, they manage 48 and 51\% of the UCTE energy and power respectively but
acummulate 85, 68 and 79\% of the UCTE average interruption time, power loss and energy not delivered, respectively.
On the contrary, though networks in group 2 (i.e., networks with $f_c^{real} > f_c^{theor}$) represent a mere 33\% 
of the whole UCTE nodes, they manage 46 and 44\% of the UCTE energy and power respectively (similar to those of group 1)
but, even so, they acummulate only 15, 32 and 21\% of the UCTE average interruption time, power loss and energy 
not delivered, respectively. This fact would suggest a positive correlation between static topological robustness and non-topological reliability measures and, as a consequence, a clear diferentiation between two classes of networks
in terms of their level of robustness.

\section{Discussion}%%%%%%%%%%%%%%%%%%%%%%%%%%%%%%%%%%%%%%

In this paper we have extended our previous work on the robustness of the European power grid under random failures  with the intentional attacks scenario. A mean field theory approach has been used in order to analitically predict the fragility of the networks against selective removal of nodes and a significant deviation from predicted values has been found for power grids with an exponent $\gamma < 1.5$. For these networks, the real critical fraction $f_c^{real}$ is higher than the theoretical one $f_c^{theor}$ for the same $\gamma$. This suggests an increased robustness for these networks compared to those with $\gamma > 1.5$. 

In order to evaluate the real existence of this two classes of networks, namely \emph{robust} and \emph{fragile}, real reliability measures from the Union for the Co-Ordination of Transport of Electricity (UCTE) have been used.  It has been found that there seems to exist indeed a positive correlation between static topological robustness measures and real non-topological reliability measures. This correlation shows that networks in the \emph{robust} class  (i.e., networks with $f_c^{real} > f_c^{theor}$), though representing only 33\% of the UCTE nodes under study and managing a similar amount of power and energy than that of the networks in the \emph{fragile} class, acummulate much less percentage of the whole UCTE average interruption time, power loss and energy not delivered.  How this can be related with the internal topological structure of the networks and the subgraphs abundances is actually a main point under study and will be explored  elsewhere. 

This feature is of obvious importance. Up to this date and as far as we know, no such correlation between topological and dynamical features has been encountered in any study related to complex networks structure and dynamics. From the power industry point of view, constanly facing the challenge of meeting growing demands with security of supply  at the lowest possible spenditure in infrastructures, the implications of this feature would permit new rather than traditional approaches to   contingency-based planning criteria \cite{Willis2004}. One of these traditional, and widely used, planning criteria is the so-called $(N-X)$ criterion. It assumes that no interruption of service can occur in a system with $N$ units of equipment due to isolation of $X$ outaged components. Without any topological feedback, the $(N-X)$ methodology (a) requires fast breaker operation to open any circuit pathway that has been faulted as well as to close the alternate path to service and (b) pushes the system to an increasing interconnection complexity as its utilization ratio (i.e., ratio between peak load and capacity of subtransmission lines and substation transformers) increases in time (aging infrastructures). Though aging infrastructures, excessive power delivered through increasing long distances and other possible causes may influence the increasing fragility of the power grids, it seems reasonable to think that, on a topological basis, the application of the $(N-X)$ contingency-based criteria, though originally intended to avoid interruptions in power service, would difficult, at the same time, the islanding of disturbances (i.e., the more connected an element is, the easier would be for a disturbance to reach). In other words: the same criteria that successfully has served to increase reliability in power systems through the late 20th century might now be responsible for the difficulties encountered in preventing perturbations, blackouts or isolating the different power grid elements.

Over the last years, and mainly due to economic imperatives, contingency-based planning criteria has been 
gradually pervaded  by reliability-based planning criteria. In the later, the prevention of likely contingencies of severe impact is  considered much more effective than that of low-probability and low impact. Nonetheless, this fact leaves the main conception of $(N-X)$ criteria still valid and at work in most of the ongoing grid's planning processes. Following the former discussion, we would suggest to add a third topology-based planning methodology, in order to take this fact into account.

\begin{acknowledgments}%%%%%%%%%%%%%%%%%%%%%%%%%%%%%%%%%
The authors thank Ricard Bosch, Iñaki Candela and Robert Neville for useful discussions. 
This work has been supported by grants FIS2004-05422 and by the Santa Fe Institute.
\end{acknowledgments}

%\bibliography{pre}  % Produces the bibliography via BibTeX.

\end{document}